\documentclass[aps,showpacs]{revtex4}
\usepackage{epsfig,graphics,amssymb,amsmath,subeqnarray}
\def\d{{\rm d}}\def\p{\partial}\def\u{{\bf u}}\def\e{{\bf e}}\def\btau{\boldsymbol{\tau}}\def\bgam{\boldsymbol{\gamma}}\def\gamd{\dot\bgam}\def\De{{\rm De}}\def\n{{\bf n}}\def\dgamd{\stackrel{\triangledown}{\gamd}}\def\bs{{\boldsymbol{\sigma}}}\def\dbtau{\stackrel{\triangledown}{\btau}}\def\x{{\bf x}}\def\U{{\bf U}}\def\F{{\bf F}}\def\L{{\bf L}}\def\O{{\boldsymbol\Omega }}\def\n{{\bf n}}\def\BS{{\boldsymbol \Sigma}}\def\H{{\cal H}}\def\G{{\cal G}}\def\A{{\cal A}}\def\B{{\cal B}}\def\M{{\bf M}}\def\N{{\bf N}}\def\H{{\bf H}}\def\r{{\bf r}}

\begin{document}

\title{Life at high Deborah number}
\author{Eric Lauga\footnote{Electronic mail: \texttt{elauga@ucsd.edu}}}
\affiliation{                    
Department of Mechanical and Aerospace Engineering, 
University of California San Diego,
9500 Gilman Dr.,  La Jolla CA 92093-0411, USA.
}



\begin{abstract}
In many biological systems, microorganisms swim through complex polymeric fluids, and usually deform the medium at a rate faster than the inverse fluid relaxation time. We address the basic properties of such life at high Deborah number  analytically by considering  the small-amplitude swimming of a body in an arbitrary complex fluid. Using asymptotic analysis and differential geometry, we show that for a given swimming gait, the time-averaged leading-order swimming kinematics of the body can be expressed as an integral equation on the solution to a series of simpler Newtonian problems.  We then use our results  to demonstrate that Purcell's scallop theorem, which states that time-reversible body motion cannot be used for locomotion in a Newtonian fluid,  breaks down in  polymeric fluid environments.

\end{abstract}

\date\today
\maketitle

\section{Introduction}


The physics of cell locomotion  in viscous fluids affects many important biological processes \cite{braybook}, such as the journey of spermatozoa through the mammalian female reproductive tract \cite{fauci06}, the mechanisms by which motile bacteria  are able to progress towards high nutrient concentration \cite{bergbook}, and the availability of  plankton  as food source for higher organisms in the ocean \cite{plankton_book}.

In many relevant instances, cells  have to move through complex  fluids, in particular during reproduction. In order to reach the uterus of the female and continue their journey towards the ovum, mammalian spermatozoa cells have to progress through the cervical mucus, a highly viscous and highly elastic cross-linked polymeric gel \cite{fauci06}. The rheology  of cervical mucus depends on its hydration \cite{tam80}, and varies during the female menstrual cycle \cite{wolf77_3}, but its typical viscosity is two to four orders magnitude larger than that of water \cite{litt76,meyer76, wolf77_1,wolf77_2}, and its  typical relaxation time, $\lambda$, is in the $1-10$~s range \cite{litt76,wolf77_2,tam80}. Since spermatozoa actuate their flagella with typical frequencies $\omega \sim 20-50$~Hz \cite{brennen77}, cell locomotion through the cervical mucus occurs therefore at high Deborah number, $\De=\lambda \omega \gg 1$, and elastic effects are expected to play a crucial role.

Building on twenty years of research on the mechanics of locomotion in simple (Newtonian) fluids, Purcell  detailed in  his 1977 classical paper  the  physical principles of life at low Reynolds number  \cite{purcell77}. In contrast, the basic properties of  life at high Deborah number are not understood. 
Calculating the swimming speed of a given organism in a given complex fluid has only been solved for infinite models  \cite{lauga07,Fu08}, and the most basic questions remain unanswered: How different are the locomotion kinematics from those obtained in a Newtonian fluid? Can the nonlinear rheological properties of the fluid (in particular shear-thinning viscosity and normal stress differences \cite{birdvol1}) be exploited to design new propulsion methods?

Here, we address the problem of locomotion at high Deborah number analytically. We show that for small-amplitude swimming of a body in an arbitrary complex fluid, the swimming kinematics can be expressed as an integral equation on the solution to a series of simpler problems (motion in a Newtonian fluid), thereby  bypassing the explicit solution  for the complete flow field. We then exploit our results  to demonstrate explicitly that Purcell's scallop theorem --- which states that time-reversible body motion cannot be used for locomotion in a Newtonian fluid \cite{purcell77} --- breaks down in a polymeric fluid.

\section{Newtonian swimming}

\begin{figure}[t]
\begin{center}
 \includegraphics[width=0.5\textwidth]{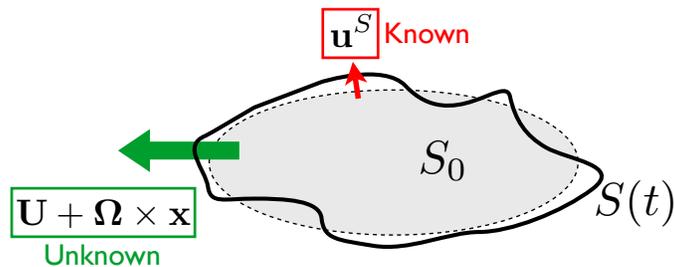}
\end{center}
 \caption{General statement of the swimming problem in a fluid: A body of fixed volume deforms its shape $S(t)$ in a time-periodic fashion around an average shape, $S_0$. The  surface deformation  is prescribed in the swimming frame (Eulerian velocity $\u^S$), and the unknown solid-body swimming kinematics (velocity, $\U$; rotation rate,  $\O$) are determined using the constraint of force-free and torque-free motion.}
\label{general}
\end{figure}

We first recall the solution to the swimming problem in a Newtonian flow  \cite{stone96}. Consider an isolated three-dimensional swimmer of instantaneous surface $S$ with normal $\n$ into the fluid.  Lorentz' reciprocal theorem   \cite{happel} states that for two arbitrary solutions of Newtonian Stokes flows with the same viscosity, $(\u,\bs)$ and $(\hat\u,\hat\bs)$, we have the equality
\begin{equation}\label{reciprocal}
\int\!\!\!\int _{S}\u\cdot \hat\bs\cdot \n \,\d S=
\int\!\!\!\int _{S}\hat\u\cdot \bs\cdot \n \,\d S,
\end{equation}
where $\u(\hat\u)$ and $\bs(\hat\bs)$ are the velocity and stress fields. For $(\u,\bs)$ we consider the swimming problem (Fig.~\ref{general}): In the swimming frame, the body prescribes its instantaneous surface velocity, $\u^S$, and as a result moves  with instantaneous (but unknown) swimming velocity $\U$ and rotation rate $\O$, so that the surface velocity is given in the lab frame by $\u  =   \U + \O\times \x^S + \u^S$, for any point $\x^S$ on its surface (in this paper, rotation rates and torques will be defined with respect to some arbitrary origin). For $(\hat\u,\hat\bs)$, we consider solid body motion of $S$ with velocity  $\hat \U$ and rotation rate $\hat\O$, so that $\hat\u  =  \hat \U + \hat\O\times \x^S$ on the surface. The body in the hat problem is therefore subject to an instantaneous force, $\hat\F=\int\!\!\!\int \hat\bs \cdot \n\,\d S$, and torque, $\hat\L=\int\!\!\!\int \x \times (\hat\bs \cdot \n)\,\d S$. Exploiting the fact that locomotion at low Reynolds numbers is force-free and torque-free, {\it i.e.}
\begin{equation}
\int\!\!\!\int _{S}\bs\cdot \n\,\d S=\int\!\!\!\int _{S}\x\times( \bs\cdot \n)\,\d S={\bf 0},
\end{equation}
Eq.~\eqref{reciprocal} leads to an equation for $\U$ and $\O$ as
\begin{equation}\label{SS}
\hat \F\cdot \U +\hat \L \cdot \O = -\int\!\!\!\int _{S}\n\cdot\hat \bs \cdot \u^S\,\d S.
\end{equation}
Eq.~\eqref{SS} states that, for a given shape ($S, \n$), and a given swimming gait ($\u^S$), all six components of the swimming kinematics, $(\U,\O)$ can be calculated using solely information about the dual problem of solid-body motion ($\hat\F$ and $\hat\L$ in Eq.~\eqref{SS} are arbitrary). Importantly, we note that the value of the fluid viscosity is irrelevant: As all hat terms in Eq.~\eqref{SS} are proportional to the viscosity, the relationship between the swimming gait ($\u^S$) and the swimming kinematics ($\U , \O$) is independent of the viscosity.

\section{Locomotion in non-Newtonian fluids}

We now consider the case where swimming occurs in a complex fluid. The stress tensor, $\bs$, includes an isotropic part (the pressure, $p$), and a deviatoric component, $\btau=\bs+p {\bf 1}$. We assume the velocity field, $\u$, to be incompressible, and therefore the equations for mechanical equilibrium in the absence of inertia are written as $\nabla p = \nabla\cdot \btau$ and $\nabla \cdot \u = 0$. 
For constitutive modeling, we assume that $\btau$ can be written as a sum of different modes, $\btau=\sum_{i}\btau^i$, where each stress $\btau^i$ satisfies a nonlinear differential constitutive relationship of the form
\begin{equation}\label{const}
(1+ \A_i ) \btau^i  +  \M_i(\btau^i ,\u) = \eta_i(1+\B_i)\gamd+\N_i(\gamd ,\u).
\end{equation}
In Eq.~\eqref{const},  $\gamd=\nabla\u + \nabla \u^T$ is the shear rate tensor, $\A_i$ and $\B_i$ are two sequences of  linear differential operators in time representing polymer  relaxation and  retardation respectively, $\M_i$ and $\N_i$ are two sequences of symmetric nonlinear operators representing transport and stretching of the polymeric microstructure by the flow, and $\eta_i$ is the zero-shear rate viscosity of the $i$-th mode. The relationship between stresses and strain rates described by Eq.~\eqref{const} is a very   general differential constitutive relationship  \cite{bird76,birdvol1,birdvol2,tanner88,larson99}, which includes as particular cases all classical models of polymeric fluids\footnote{It includes in particular: Second and n-th order fluid, all Oldroyd-like models (upper-convected Maxwell, lower-convected Maxwell, corotational Maxwell, Oldroyd-A, Oldroyd-B, corotational Oldroyd, Oldroyd 8-constant model,  Johnson-Segalman-Oldroyd),  the Giesekus and Phan-Thien-Tanner models,  Generalized Newtonian fluids, and all  multi-mode version of these models. Furthermore, although  FENE-P is only exactly in this form, it becomes in the asymptotic limit of small surface deformation \cite{lauga07}, so the relationship is also valid for FENE-P and FENE-P-like models.}.

We consider a body performing periodic small-amplitude swimming motion in a fluid described by Eq.~\eqref{const}. 
Its undeformed surface shape is termed $S_0$, parameterized by $\x_0^S$, and we define $\epsilon$ as  the 
 amplitude of the periodic surface distortion non-dimensionalized by a typical swimmer length ($\epsilon\ll1$).  Material points on the swimmer shape, $\x^S$, are assumed to display time-variations of the form $\x^S(\x^S_0,t) = \x^S_0 + \epsilon \x^S_1 (\x^S_0,t)$,
and the function $\x^S_1$ is assumed to be periodic in time with period $T$.
Such Lagrangian boundary motion forces the fluid to move through the no-slip boundary condition,  $\u^S(\x^S)=\p \x^S/\p t$.

We solve the swimming problem as a  domain perturbation expansion, where the fields of interest  are written as regular perturbation expansions, with   
boundary conditions expressed on $S_0$. Specifically, we write 
\begin{equation}
\{\mathbf{u},\btau,p,\bs\}= \epsilon\{\mathbf{u}_{1} ,{\btau}_{1},p_1, \bs_1 \} 
+ \epsilon^2\{\mathbf{u}_{2},{\btau}_{2},p_2, \bs_2 \}+...
\end{equation}
which are all functions of $(\x,t)$, and are defined on the zeroth-order surface $S_0$. The boundary condition for the  surface velocity reads
\begin{equation}
\u^S= \epsilon \u^S_1 (\x^S_0,t) + \epsilon^2 \u^S_2 (\x^S_0,t)+...
\end{equation}
and swimming occurs with the kinematics
\begin{equation}
\{\U,\O\}= \epsilon\{\U_1,\O_1\}+ \epsilon^2\{\U_2,\O_2\}+...
\end{equation}
so that  on the swimmer  surface we have $\u_n  =   \U_n + \O_n\times \x_0^S + \u_n^S$,  for $n=1,2,...$. Based on the Newtonian case, we expect to obtain no swimming at order $\epsilon$, but non-zero time-averaged locomotion at order $\epsilon^2$ \cite{taylor51}.

\subsection{First-order solution}

At order $\epsilon$, the constitutive model, Eq.~\eqref{const}, is linearized 
\begin{equation}\label{order1_i}
(1+ \A_i ) \btau^i_1  = \eta_i(1+\B_i)\gamd_1,
\end{equation}
associated with boundary conditions  $\u^S_1=\p \x^S_1/\p t$, evaluated at $(\x^S_0,t)$. Since the surface motion is time-periodic, we introduce Fourier series, and write, for any field $f(t)$,  $f(t)=\sum_{-\infty}^\infty \tilde f ^{(n)}e^{i n \omega t}$ where $\omega = 2\pi/T$ and 
$\tilde f^{(n)}= \frac{1}{T}\int_0^T f (t)e^{-i n \omega t}\,\d t$.  In Fourier space, Eq.~\eqref{order1_i} then becomes 
\begin{equation}
\tilde \btau^{i,(n)}_1 (\x) = \G_i(n) \tilde\gamd_1^{(n)}(\x),
\end{equation}
where $\G_i(n)$ is the $i^{\rm th}$ relaxation modulus  of the  $n^{\rm th}$ Fourier mode. Since we have $\btau=\sum_{i}\btau^i$, we get the constitutive equation for the total first order deviatoric stress as
\begin{equation}\label{order1_fourier}
\tilde \btau_1^{(n)} (\x) = \G(n) \tilde\gamd_1^{(n)}(\x), \quad \G(n)=\sum_i\G_i(n).
\end{equation}
We see from Eq.~\eqref{order1_fourier} that, for each Fourier mode, the swimming problem is a Newtonian problem with a complex viscosity ($\G$). We have to solve $\nabla \tilde p_1^{(n)} = \G(n)  \nabla^2 \tilde \u_1^{(n)}$, $\nabla \cdot \tilde\u_1^{(n)} = 0$, subject to the boundary condition $\tilde \u_1^{(n)} (\x_0^S)=
 \tilde \U_1^{(n)} +  \tilde \O_1^{(n)} \times \x_0^S + \tilde \u^{S,(n)}_1(\x_0^S)$. 
Applying  Eq.~\eqref{SS}, we obtain the swimming kinematics for each Fourier mode
\begin{equation}\label{order1_swim_fourier}
\hat \F\cdot \tilde \U_1^{(n)} +\hat \L \cdot \tilde\O_1^{(n)} = -\int\!\!\!\int _{S_0}\n_0\cdot\hat \bs \cdot \tilde\u^{S,(n)}_1(\x_0^S)\,\d S.
\end{equation}
Since the value of the viscosity for the hat fields in Eq.~\eqref{order1_swim_fourier} is arbitrary, we can take it to be some fixed reference viscosity. In addition, as $S_0$ does not depend on time, we can Fourier-invert Eq.~\eqref{order1_swim_fourier} to obtain the locomotion in the time domain
\begin{equation}\label{order1_final}
\hat \F\cdot \U_1(t) +\hat \L \cdot \O_1(t) = -\int\!\!\!\int _{S_0}\n_0\cdot\hat \bs \cdot \u^S_1(\x_0^S,t)\,\d S.
\end{equation}
The solution at order $\epsilon$ leads thus to the same swimming kinematics as in a Newtonian flow (Eq.~\ref{SS}). In addition, since  $\u^S_1=\p \x^S_1/\p t$, we get that $\langle \u_1^S \rangle =0$, where $\langle . \rangle$ denotes time-averaging over one period of body deformation ({\it i.e.} the zeroth Fourier mode). From Eq.~\eqref{order1_final} we therefore see that $\langle \U_1\rangle~=~\langle \O_1 \rangle~=~{\bf 0}$. As in the Newtonian case, there is no time-averaged locomotion at leading order, and swimming is quadratic in the amplitude of the surface motion \cite{taylor51}.

\subsection{Second-order solution}

At order $\epsilon^2$, the constitutive relationship for each mode, Eq.~\eqref{const}, is written as 
\begin{equation}\label{taui_order2}
(1+ \A_i ) \btau^i_2   = \eta_i(1+\B_i)\gamd_2+\H_i[\u_1],
\end{equation}
with
\begin{equation}\label{Hi}
\H_i=\u_1\cdot
\{
\gamd_1:
\left[(\nabla_{\gamd} \nabla_\u 
\N_i)\right]
-\btau_1^i: 
\left[(\nabla_{\btau^i} \nabla_\u 
\M_i)\right]\}
\end{equation}
where the gradients in Eq.~\eqref{Hi} are evaluated at $(\bf 0,\bf 0)$, and with  $\btau^i_1$ and $\gamd_1$ related through Eq.~\eqref{order1_i}.
Since we are interested in the time-averaged swimming motion, which we expect occurs at $O(\epsilon^2)$, 
we now consider  only time-averaged quantities. Averaging Eq.~\eqref{taui_order2} leads to
\begin{equation}
\langle  \btau^i_2 \rangle   = \eta_i\langle \gamd_2 \rangle +\langle \H_i[\u_1]\rangle ,
\end{equation}
and therefore the time-averaged stress is given by
\begin{equation}\label{CR_order2}
\langle \bs_2 \rangle = -\langle p_2 \rangle {\bf 1}+ \eta \langle \gamd_2 \rangle+\langle \BS[\u_1]\rangle,
\end{equation}
where $\eta=\sum_i\eta_i$ and $\langle \BS[\u_1]\rangle=\sum_i\langle \H_i[\u_1]\rangle $. 

To derive the swimming kinematics, we apply the principle of virtual work using the following two problems: (i) solid body motion of the shape $S_0$  in a Newtonian fluid of viscosity $\eta$ (the same viscosity as in Eq.~\ref{CR_order2})
, with velocity and stress fields given by $\hat\u$ and $\hat\bs$ and (ii) time-averaged swimming with flow velocity 
$\langle \u_2\rangle $ and stress field $\langle\bs_2\rangle$ given by Eq.~\eqref{CR_order2}.

Since mechanical equilibrium is written $\nabla \cdot \hat\bs= \nabla\cdot \langle\bs_2\rangle={\bf 0}$, we have equality of  their dot products with the opposite velocity field, $[\nabla \cdot \hat\bs]\cdot \langle \u_2\rangle=  [\nabla\cdot \langle\bs_2\rangle]\cdot \hat\u$, and integration over the volume of fluid $V_0$ outside of $S_0$  leads to
\begin{eqnarray}\label{order2}
&&\int\!\!\!\int _{S_0}\n_0\cdot \hat\bs\cdot \langle \u_2\rangle\,\d S
-
\int\!\!\!\int _{S_0}\n_0\cdot \langle  \bs_2\rangle\cdot \hat\u\,\d S\\
&&=
\int\!\!\!\int\!\!\!\int_{V_0} \langle \bs_2\rangle:\nabla \hat\u \, \d V
-
\int\!\!\!\int\!\!\!\int_{V_0} \hat\bs:\nabla \langle \u_2\rangle  \, \d V\nonumber,
\end{eqnarray}
where we have used integration by parts, and the fact that $\n_0$ is directed into the fluid.
If we then insert  Eq.~\eqref{CR_order2} into the right-hand side of Eq.~\eqref{order2} we obtain
\begin{eqnarray}
&&\int\!\!\!\int\!\!\!\int_{V_0} \langle \bs_2\rangle:\nabla \hat\u \, \d V
-
\int\!\!\!\int\!\!\!\int_{V_0} \hat\bs:\nabla \langle \u_2\rangle  \, \d V\\
&& =\int\!\!\!\int\!\!\!\int_{V_0}\langle \BS[\u_1]\rangle : \nabla \hat\u \, \d V\nonumber,
\end{eqnarray}
and the Newtonian components of  both $\hat\bs$ and $\langle \bs_2\rangle $ have disappeared due to symmetry and incompressibility. Consequently,  Eq.~\eqref{order2}  becomes
\begin{eqnarray}\label{order2_better}
&&\int\!\!\!\int _{S_0}\n_0\cdot \hat\bs\cdot \langle \u_2\rangle\,\d S
-
\int\!\!\!\int _{S_0}\n_0\cdot \langle  \bs_2\rangle\cdot \hat\u\,\d S\\
&&=\int\!\!\!\int\!\!\!\int_{V_0}\langle \BS[\u_1]\rangle : \nabla \hat\u \, \d V,\nonumber
\end{eqnarray}
and only the deviation from Newtonian behavior, $\BS$, remains in the integral formula. This result is reminiscent of  past work quantifying small viscoelastic effects on particle motions~\cite{leal80}. 

On the surface $S_0$ we have $\langle\u_2 \rangle=\langle\U_2\rangle + \langle\O_2\rangle\times \x_0^S + \langle \u_2^S\rangle$, where  
a Taylor expansion of the boundary conditions around $\x^S_0$ leads to 
$\u_2^S(\x^S_0,t)=-\x_1^S \cdot \nabla \u_1$,  so that Eq.~\eqref{order2_better} becomes
\begin{eqnarray}\label{order2_almost}
&&\hat \F\cdot \langle \U_2 \rangle  +\hat \L \cdot \langle  \O_2 \rangle=
- \int\!\!\!\int _{S_0}\n_0\cdot\hat \bs \cdot \langle \u_2^S\rangle \,\d S
\\ \nonumber
&&+\int\!\!\!\int _{S_0}\n_0\cdot \langle  \bs_2\rangle\cdot \hat\u\,\d S
+\int\!\!\!\int\!\!\!\int_{V_0}\langle \BS[\u_1]\rangle : \nabla \hat\u \, \d V.
\quad\quad \end{eqnarray}

The final step in the calculation consists in enforcing the force-free and torque-free condition for the swimmer. On $S_0$ we have $\hat\u = \hat\U + \hat\O\times \x_0^S$, so that
\begin{eqnarray}\label{FO2}
\int\!\!\!\int _{S_0}\n_0\cdot \langle  \bs_2\rangle\cdot \hat\u\,\d S&=&
\left[\int\!\!\!\int _{S_0}\n_0\cdot \langle  \bs_2\rangle\,\d S\right] \cdot \hat\U\,\,\,
\\
&+&
\left[\int\!\!\!\int _{S_0}\x_0^S\times (\n_0\cdot \langle  \bs_2\rangle)\,\d S\right]\cdot \hat\O\nonumber.
\end{eqnarray}

The terms in brackets in Eq.~\eqref{FO2} are related to the second forces and torque on the swimmer at order $\epsilon^2$, and can be evaluated using differential geometry. Let us write the time-varying shape of the swimmer as $\x^S = \x^S_0 + \epsilon \n_0 \delta_1 (\x^S_0,t)+...$, where 
the function  $\delta_1$, with units of length,  represents the normal extent of the surface deformations. When  $\delta_1=0$, the shape of the swimmer does not change with time, and all surface motion is tangential ($\u^S\cdot\n=0$, so-called squirming motion), whereas for $\delta_1\neq 0$ the body also undergoes normal deformation and varies its  shape periodically. If we write the normal to the surface as $\n=\n_0 + \epsilon \n_1+...$,  differential geometry considerations leads to the evaluation of the force, $\F_2 $, and torque, $\O_2$, on the swimmer at order $\epsilon^2$, as given by
\begin{subeqnarray}\label{F02_total}
\F_2  =  
\int\!\!\!\int _{S_0}
\left[\n_1\cdot \bs_1 +  \n_0\cdot\left( \bs_2 +  \delta_1  \frac{\p \bs_1}{\p n}\right)\right]\,\d S, \quad \quad \,\, \,
\\
\O_2  =  
\int\!\!\!\int _{S_0}
\x_0^S\times 
\left[\n_1\cdot \bs_1 +  \n_0\cdot\left( \bs_2 +  \delta_1  \frac{\p \bs_1}{\p n}\right)\right]\, \d S\,\, \,
\end{subeqnarray}
where $\p/\p n\equiv\n_0\cdot\nabla$ denotes the normal derivative to $S_0$.
Since locomotion occurs with no force or torque, we have $\F_2={\O_2}={\bf 0}$, and therefore after taking time-averages of Eq.~\eqref{F02_total}, we obtain 
\begin{subeqnarray}\label{diffgeom}
&& \int\!\!\!\int _{S_0}\n_0\cdot \langle  \bs_2\rangle\,\d S \\ && =- \int\!\!\!\int _{S_0}
\left[  \langle \n_1\cdot \bs_1 \rangle  + \n_0\cdot \left\langle \delta_1 \frac{\p \bs_1}{\p n}\right\rangle
\right]\,\d S, \nonumber \\
&& \int\!\!\!\int _{S_0}\x_0^S\times (\n_0\cdot \langle  \bs_2\rangle)\,\d S\\
&&=-
\int\!\!\!\int _{S_0}\x_0^S\times 
\left[  \langle \n_1\cdot \bs_1 \rangle  + \n_0\cdot \left\langle \delta_1 \frac{\p \bs_1}{\p n}\right\rangle
\right]\,\d S\nonumber.
\end{subeqnarray}

\subsection{Life at high Deborah number}
To obtain the final integral formula, we insert the result of Eq.~\eqref{diffgeom} into Eqs.~\eqref{order2_almost} and \eqref{FO2} to obtain the integral relationship
\begin{eqnarray}\label{final}
&& \hat \F\cdot\langle \U_2 \rangle+\hat \L \cdot \langle\O_2\rangle = \\\nonumber
& &- \int\!\!\!\int _{S_0}\n_0\cdot\hat \bs \cdot \langle \u_2^S\rangle \,\d S 
+\int\!\!\!\int\!\!\!\int_{V_0}\langle \BS[\u_1]\rangle : \nabla \hat\u \, \d V\\\nonumber
&& -
\left\{
\int\!\!\!\int _{S_0}
\left[  \langle \n_1\cdot \bs_1 \rangle  + \n_0\cdot \left\langle \delta_1 \frac{\p \bs_1}{\p n}\right\rangle
\right]\,\d S\right\}\cdot\hat\U
\\\nonumber
&&-\left\{
\int\!\!\!\int _{S_0}\x_0^S\times 
\left[  \langle \n_1\cdot \bs_1 \rangle  + \n_0\cdot \left\langle \delta_1 \frac{\p \bs_1}{\p n}\right\rangle
\right]\,\d S\right\}\nonumber\cdot\hat\O.
\end{eqnarray}
The result expressed by Eq.~\eqref{final} is the non-Newtonian equivalent of the Newtonian integral 
formula, Eq.~\eqref{SS}. It shows that one can compute the time-averaged swimming kinematics for locomotion in a complex fluid, 
using knowledge of a series of simpler problems. 
Indeed, to compute $\langle \U_2 \rangle$ and $\langle \O_2 \rangle$ from Eq.~\eqref{final}, and beyond the necessary 
knowledge of the surface motion of the swimmer, one needs to know the velocity and stress field 
for solid body motion of $S_0$ ({\it i.e.} the fields $\hat \u$ and $\hat \bs $), and the velocity and stress 
field for the first-order solution ({\it i.e.} $ \u_1$ and $ \bs_1$). As discussed above, 
and shown in Eq.~\eqref{order1_fourier}, the  first order solution can be found in frequency space by solving a 
series of Newtonian flow problems. Consequently, the computational complexity to evaluate the terms in 
Eq.~\eqref{final} is that of  a succession of Newtonian flow problems, and therefore using this method one bypasses entirely
the calculation of the second-order flow and stress field. Notably, the final result can be applied to flows with arbitrary large Deborah 
numbers, as is relevant in cell locomotion. Note also that for squirming motion of the sphere, for which $\delta_1=0$ and $\n_1={\bf 0}$,  Eq.~\eqref{final} is greatly simplified. 


\section{Breakdown of the scallop theorem}
As an application of our results, we demonstrate that Purcell's scallop theorem \cite{purcell77} breaks down in a polymeric fluid. We consider the axisymmetric squirming motion of a sphere (radius, $a$) in an Oldroyd-B fluid \cite{bird76,birdvol1,birdvol2,tanner88,larson99}. 
Purcell's scallop theorem states that if the surface motion is time-reversible, we have $\langle \U \rangle = \langle \O \rangle ={\bf 0}$
and therefore the Newtonian contribution to Eq.~\eqref{final} averages to zero,
\begin{equation}
 \int\!\!\!\int _{S_0}\n_0\cdot\hat \bs \cdot \langle \u_2^S\rangle \,\d S={\bf 0}.
\end{equation}
In addition, we consider axisymmetric surface deformation  so that we have $\langle \O_2 \rangle ={\bf 0}$. As a consequence,  the integral equation  leading the average swimming speed, Eq.~\eqref{final},   simplifies to
\begin{equation}\label{tosolve}
\hat \F\cdot \langle \U_2\rangle  =
\int\!\!\!\int\!\!\!\int_{V_0}\langle \BS[\u_1]\rangle : \nabla \hat\u \, \d V.
\end{equation}
For constitutive modeling, we consider an Oldroyd-B fluid,  which represents a polymeric fluid as a dilute solution of elastic dumbbells  \cite{larson99}, and for which  the relationship between stresses and rate of strains is given by
\begin{equation}\label{oldroydB}
\btau +\lambda_1 \dbtau= \eta [\gamd + \lambda_2 \dgamd ],
\end{equation}
where $\stackrel{\triangledown}{{\bf a}} ={\p {\bf a}}/{\p t} + {\bf u}\cdot \nabla{\bf a} - (\nabla \u^T\cdot {\bf a} + {\bf a}\cdot \nabla \u)$ is the upper-convected derivative for the tensor $\bf a$. In Eq.~\eqref{oldroydB}, $\lambda_1 $ and $\lambda_2$ are, respectively, the relaxation and  retardation time scales for the fluid. If $\eta_s$ denotes the solvent viscosity, and $\eta$ the polymer viscosity, we have $\lambda_2/\lambda_1=\eta_s/\eta < 1$.

\begin{figure}[t]
\begin{center}
 \includegraphics[width=0.6\textwidth]{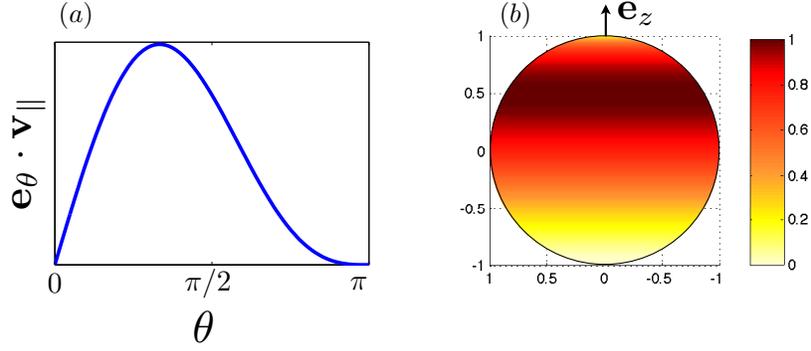}
\end{center}
 \caption{Distribution of surface velocity,  ${\bf v}_\parallel $, (arbitrary units) for axisymmetric spherical squirming motion and breakdown of the scallop theorem. (a): Surface velocity as a function of the polar angle, $\theta$;   (b):  Color map of the the surface velocity. Both figures illustrate the fore-aft asymmetry of the tangential surface motion.}
\label{squirming}
\end{figure}

For a  time-reversible deformation, we consider a simple sinusoidal gait of the  form $\u_1^S(\x_0^S,t)= {\bf v}_\parallel^S(\x_0^S)\cos\omega t$, so that $\tilde \u_1^{S,(n)}(\x_0^S)={\bf 0}$ for all $n\neq \pm1$, and $\tilde \u_1^{S,(\pm1)}(\x_0^S)={\bf v}_\parallel^S(\x_0^S)/2$ otherwise. At order $\epsilon$, only the Fourier modes with $n\neq \pm1$ are non-zero, and we have 
\begin{equation}
\tilde \btau_1^{(1)} =  {\cal G} \tilde \bgam_1^{(1)}
,\quad \tilde \btau_1^{(-1)} =  {\cal G}^* \tilde \bgam_1^{(-1)}
,\quad {\cal G}=\eta\frac{1+i\lambda_2\omega}{1+i\lambda_1\omega},
\end{equation}
where $\{.\}^*$ denotes the complex conjugate. The spatial distribution of surface deformation, described by ${\bf v}_\parallel$, is assumed to be axisymmetric. The symmetry axis is denoted $\e_z$ (see Fig.~\ref{squirming}), the polar angle  is $\theta$, and is associated with the orientation vector  $\e_\theta$. In the frame moving with the swimmer, we prescribe  
\begin{equation}\label{vp}
{\bf v}_\parallel^S (\x^S_0)= 3 a\omega \sin\theta(1+\cos\theta)\, \e_\theta, 
\end{equation}
which is  illustrated in Fig.~\ref{squirming}. Note that the velocity distribution described by Eq.~\eqref{vp} is  fore-aft asymmetric, which is  necessary in order to obtain net locomotion with an  actuation varying sinusoidally in time.

Given Eq.~\eqref{vp}, we can then calculate the unsteady swimming at order $\epsilon$ from Eq.~\eqref{order1_final}, and we find $\U_1= 2 a \omega \cos\omega t \e_z$. As a result, the surface distribution of velocity in the lab frame is given by 
$\u_1(\x_0^S,t)= {\bf v}_\parallel(\x_0^S)\cos\omega t$, where
\begin{equation}\label{vp_lab}
{\bf v}_\parallel(\x^S_0)= 2 a\omega\cos\theta\, \e_r +  a\omega \sin\theta(1+3\cos\theta)\,\e_\theta.
\end{equation}
Given  Eq.~\eqref{order1_swim_fourier}, it is then easy to show that each Fourier component of the entire flow field  is identical to that obtained in the Newtonian problem. Consequently, if  ${\bf v}_\parallel(\x)$ denotes the Newtonian velocity field associated with the lab-frame boundary conditions  ${\bf v}_\parallel(\x_0^S)$ on $S_0$, we obtain at first order $\u_1(\x,t) = {\bf v}_\parallel(\x)\cos\omega t$. The velocity field ${\bf v}_\parallel(\x)$ with boundary conditions from Eq.~\eqref{vp_lab} can be found using the Legendre polynomials method pioneered by Blake \cite{blake71a}, and we get
${\bf v}_\parallel ={v}_{\parallel,r} \e_r +{v}_{\parallel,\theta}\e_\theta$ with
\begin{subeqnarray}\label{vp_labframe}
{v}_{\parallel,r}
=  a\omega \left[ 2  \frac{a^3}{r^3}\cos\theta +    \frac{3}{2}(3\cos^2\theta-1)\left(\frac{a^4}{r^4}-\frac{a^2}{r^2}\right)\right],\,\\
{v}_{\parallel,\theta}
=   a\omega \left[ \frac{a^3}{r^3}  \sin\theta +  3 \frac{a^4}{r^4}\sin\theta\cos\theta \right]\cdot
\quad
\quad
\quad
\quad
\quad
\quad\,\,\,\,
\end{subeqnarray}

At order $\epsilon^2$, straightforward algebra allows us to obtain the deviation from Newtonian behavior, in Eq.~\eqref{CR_order2},   as \cite{lauga07}
\begin{eqnarray}\label{Sigma}
\langle\BS[\u_1]\rangle&=&
\frac{\eta (\lambda_2-\lambda_1)}{2(1+\text{De}^2)} 
\times \\
&&  \left[
 {\bf v}_\parallel\cdot\nabla\gamd_\parallel
 -\left(
\nabla{\bf v}_\parallel^T\cdot\gamd_\parallel
 +\gamd_\parallel\cdot\nabla{\bf v}_\parallel
  \right)
 \right]\nonumber
\end{eqnarray}
where $\De=\lambda_1 \omega$ is the Deborah number for the flow.

Finally, the hat problem in Eq.~\eqref{tosolve} is the  solid body translation of the sphere, with 
velocity field given by \cite{kimbook}
\begin{equation}\label{Stokes}
\hat\u= \frac{3}{4}a\left[\frac{\bf 1}{r}+\frac{\r\r}{r^3}\right] \cdot \hat \U+ \frac{1}{4}a^3 \left[\frac{\bf 1}{r^3}-\frac{3\r\r}{r^5}\right]\cdot \hat \U
\end{equation}
together with Stokes law,  $\hat\F=-6\pi\eta a \hat  \U$. 

By symmetry, we expect that average swimming will occur along the $z$ direction, so that $\langle \U_2\rangle= \langle U_2\rangle \e_z$ and by choosing  $\hat  \U= \hat U \e_z$, the left-hand side of  Eq.~\eqref{tosolve} is given by $-6\pi \eta a \hat U \langle U_2\rangle $. Given Eqs.~\eqref{Sigma}, \eqref{vp_labframe} and \eqref{Stokes}, we can evaluate the right hand side of Eq.~\eqref{tosolve} and obtain 
\begin{equation}
\int\!\!\!\int\!\!\!\int_{V_0}\langle \BS[\u_1]\rangle : \nabla \hat\u \, \d V=a^2\omega^2 \hat U \frac{\eta (\lambda_1-\lambda_2)}{1+\De^2} \frac{299\pi}{25} 
\end{equation}
Recalling that $\lambda_2 = \lambda_1\eta_s/\eta$, we obtain the explicit formula for the time-averaged swimming speed, $\langle U_2\rangle$,  of the squirming sphere as 
\begin{equation}\label{noscallop}
\langle U_2\rangle =  a\omega   \frac{ \De}{1+\De^2}\left(\frac{\eta_s}{\eta}-1\right) \Lambda,
\end{equation}
where $\Lambda = 299/150\approx 1.993$.
The result of Eq.~\eqref{noscallop} demonstrates explicitly that the scallop theorem breaks down in an Oldroyd-B fluid: The swimming gait is a sinusoidal function, and therefore time-reversible, yet the force-free body swims on average. In the Newtonian limit where $\De=0$, we have $\langle U_2\rangle =0$ and the result of the scallop theorem is recovered. 
Note that since $\eta_s < \eta$, we have $\langle U_2\rangle < 0$.  High surface shear is localized on the top of the sphere (see Fig.~\ref{squirming}b),  so this is also where high normal-stresses differences are localized, and the sphere is  being pushed from the top to swim in the $-z$ direction.


\section{Perspective}

In this paper, we have addressed the most basic problem in the  locomotion of microorganisms: For a given swimming gait, at which speed is the organism expected to swim? The solution to this problem is known in the case where the fluid is Newtonian, and given by Eq.~\eqref{SS},  but is not known for complex polymeric fluids displaying a nonlinear relationship between stress and strain rates. We have considered the time-periodic small-amplitude locomotion of a deformable body in an arbitrary complex fluid. We have shown  that the time-averaged swimming kinematics of the body (translation and rotation) are given by an integral formula on a series of simpler Newtonian problems. The final formula,  Eq.~\eqref{final}, can be applied for high Deborah numbers, which is the relevant limit for the locomotion of swimming cells in mucus, and provides the first formal framework to address locomotion in complex fluids. In addition,  our results are valid beyond the biological realm, and can be used in particular to the quantify the  locomotion of synthetic micro-swimmers  \cite{dreyfus05}.

As an application of our results we have constructed an explicit example of a deformable body that swims using a time-reversible stroke in a polymeric fluid. This example demonstrates  formally the breakdown of Purcell's scallop theorem in complex fluids for a finite-size, force-free and torque-free swimmer. Note that the final formula for the time-averaged swimming speed of the body, Eq.~\eqref{noscallop}, is reminiscent of recent work on the force generated by flapping motion in polymeric fluids \cite{normand08}. The  implication of this result, more generally, is that it is possible to exploit nonlinear rheological mechanisms (in our case, the existence of normal-stress differences) to design new swimming methods.

Finally, we note that recent work on infinite models for swimmers deforming in a wave-like fashion showed that, for a given swimming gait,  swimming is always slower in a polymeric fluid than in the Newtonian limit \cite{lauga07,Fu08}. The final integral formula for the swimming speed we obtain here, Eq.~\eqref{final}, explicitly shows that in general the beneficial vs. detrimental impact of the polymeric stresses on the swimming performance cannot be established a priori.  

The results above  could be extended in many different ways. In particular, the method of expansion outlined in the paper could be further continued, and all Fourier components of the flow at higher order in the amplitude of the surface deformation could be formally calculated. Similar work could also be performed near boundaries, or in the presence of other swimmers, and therefore could be exploited to characterize the effect of polymeric stresses on collective locomotion.
The application of our results to different swimmer geometries and various modes of surface swimming, including flagella-based, will be reported in future work.

\section{Acknowledgments}
Contributions by  Thibaud Normand and 
funding by the US National Science Foundation (grants CTS-0624830 and CBET-0746285) are gratefully acknowledged.

\bibliographystyle{unsrt}
\bibliography{NN}
\end{document}